\documentclass[journal,draftcls,onecolumn,12pt,twoside]{IEEEtran}

	\usepackage{cite}

	\usepackage[cmex10]{amsmath}

	\usepackage{array}
	\usepackage{mdwmath}
	\usepackage{mdwtab}
	\usepackage{mathtools}

	\usepackage{url}
	
	\usepackage{graphics} 
	\usepackage{epsfig} 
	\usepackage{amssymb}  
	\usepackage{hyperref} 
	\usepackage{mathtools}
	\usepackage{subfig}
	\usepackage{verbatim}
	\usepackage{color}
	\usepackage{bbm}
	\usepackage{multirow} 

	\newtheorem{theorem}{Theorem}
	\newtheorem{definition}{Definition}
	\newtheorem{lemma}{Lemma}

\newcommand{\eq}[1]{\begin{align}#1\end{align}}
\newcommand{\seq}[1]{\begin{subequations}#1\end{subequations}}

\newcommand{\E}{\mathbb{E}}
\newcommand{\cX}{\mathcal{X}}
\newcommand{\cA}{\mathcal{A}}
\newcommand{\cS}{\mathcal{S}}

\newcommand{\cP}{\Delta}
\newcommand{\tsigma}{\tilde{\sigma}}
\newcommand{\cH}{\mathcal{H}}

\newcommand{\cN}{\mathcal{N}}
\newcommand{\tgamma}{\tilde{\gamma}}

\newcommand{\defeq}{\buildrel\triangle\over =}

\newcommand{\pushright}[1]{\ifmeasuring@ #1 \else\omit\hfill$\displaystyle#1$\fi\ignorespaces}
\newcommand{\pushleft}[1]{\ifmeasuring@ #1 \else\omit$\displaystyle#1$\hfill\fi\ignorespaces}

\begin{document}
	%
	\title{ Dynamic information design}
	%
	%
	%
	\author{Deepanshu~Vasal 
	\thanks{
	e-mail: {dvasal@umich.edu}.}%
       }%
		
	\markboth{Technical note}%
	{Technical note}
	
	%

\maketitle
\begin{abstract}
We consider the problem of dynamic information design with one sender and one receiver where the sender observers a private state of the system and takes an action to send a signal based on its observation to a receiver. Based on this signal, the receiver takes an action that determines rewards for both the sender and the receiver and controls the state of the system. In this technical note, we show that this problem can be considered as a problem of dynamic game of asymmetric information and its perfect Bayesian equilibrium (PBE) and Stackelberg equilibrium (SE) can be analyzed using the algorithms presented in~\cite{VaSiAn19, Va20} by the same author (among others). We then extend this model when there is one sender and multiple receivers and provide algorithms to compute a class of equilibria of this game.
\end{abstract}

\section{Introduction}
Game theory is a powerful tool to analyze behavior among strategic agents. An engineering side of game theory is mechanism design which aims to design systems such that when played on by strategic agents who optimize their individual objectives, they achieve the same objective as envisioned by the designer. A classic and one of the most widely used practical example of mechanism design is auctions~\cite{My81} where an auctioneer asks for bids by the bidders on a private good. The auction is designed in such a way that when the strategic bidders bid on the value to maximize their own valuations, it maximizes the returns of the auctioneer. There is a huge and growing literature on the theory of Mechanism Design as well as its real world applications~\cite{Ja14}.

Information design is a relatively new field related to the field of mechanism design introduced by Kamenica and Gentzkow in~\cite{KaGe11} where a sender (designer) observes a state of the world not observed by the receiver. The sender sends a signal to the receiver about this state based on which the receiver takes an action, which determines individual rewards for both the sender and the receiver. The sender has to choose a signal that maximizes its reward. The receiver interprets the state of the world from the sender's signal knowing that the sender would have chosen a signal that maximizes its reward, and thus takes an action that maximizes its own reward. There are two notions of equilibrium that can be considered in this setting: (a) Nash equilibrium and (b) Stackelberg equilibrium.

Nash equilibrium is defined as a set of strategies of the player such that no user wants to unilaterally deviate. Thus it can be defined as a fixed-point of best responses to each player's strategies. In a stackelberg equilibrium there is a leader and a follower (sender and receiver in this case, respectively). The leader \textit{commits} to a policy that is known and observed by the receiver. Then theStackelberg equilibrium is defined as set of strategies of the players such that the receiver plays best response to the leader's committed strategy and the leader, knowing that the sender will play a best response, plays a strategy that maximizes its reward.

Since~\cite{KaGe11}, there have been a growing number of works on information design including dynamic information design where the state of the world evolves in a dynamic fashion and both sender and the receiver play a sequential game~\cite{costly2015,Tamura2012,GeKa17,multiple-senders2018,Bergemann-multiagent,Bergemann2016,heterogeneous-priors2016,Kamenica-survey,Farokhi2017,Lingenbrink,Ely-Beep,RENAULT-dynamic,Best1,Best2,HONRYO-dynamic,Skrzypacz-dynamic,Skrzypacz-selling-info,Ely-goalpost,Ely-sequential-info-design,Basu2017Dynamic,Hung-dynamic,Farhadi2018,Hamid-Allerton,Asu2020}. Authors in~\cite{Lingenbrink,Ely-Beep,RENAULT-dynamic,Best1,Best2} considered a dynamic version of the model consiodered by~\cite{KaGe11} where the state evolves as a Markov process, the sender is forward looking, however, the receiver is myopic. Recently, Farhadi and Teneketzis in~\cite{FaTe20} considered a dynamic model with evolving Markovian state and presented (Stackelberg) equilibrium strategies of the sender and the receiver, where both the sender and the receiver are fully rational. We refer the readers to~\cite{FaTe20} for an excellent introduction of information design problems.

In this note, we consider a general discrete time finite horizon model where there is a state of the system that is evolving as a controlled Markov process which is privately observed by the sender. In each time $t$, the sender sends a signal to the receiver based on which both the sender and the receiver get individual instantaneous rewards. The objective of the sender is to maximize its total expected reward over the time horizon $T$ and the objective of the receiver is to maximize its own. Thus it can be posed as a dynamic game of asymmetric information where players play alternatively. We assume both the sender and the receiver are fully rational and forward looking. We consider two equilibrium concepts:(1) Perfect Bayesian Equilibrium (PBE) which can be thought of as extension of Nash equilibrium for dynamic games of incomplete information, (2) perfect Stackelberg equilibrium (PSE) where the sender has a commitment power and commits to a policy. In this technical note, we show that game fits within the framework of the models used by the authors in~\cite{VaSiAn19,Va20}, which provides a tool to analyze Markovian Perfect Bayesian equilibrium (PBE) and Markovian Perfect Stackelberg Equilibrium (PSE) of this game. We further extend this model such that instead of one, there are multiple receivers taking actions. Based on~\cite{VaSiAn19, Va20}, we provide an algorithm to analyze PBEs of this game.

\subsection{Notation}
We use uppercase letters for random variables and lowercase for their realizations. For any variable, subscripts represent time indices and superscripts represent player indices. We use notation $ -i$ to represent all players other than player $i$ i.e. $ -i = \{1,2, \ldots i-1, i+1, \ldots, N \}$. We use notation $A_{t:t'}$ to represent the vector $(A_t, A_{t+1}, \ldots A_{t'})$ when $t'\geq t$ or an empty vector if $t'< t$. We use $A_t^{-i}$ to mean $(A^1_t, A^2_{t}, \ldots, A_t^{i-1}, A_t^{i+1} \ldots, A^N_{t})$ . We remove superscripts or subscripts if we want to represent the whole vector, for example $ A_t$  represents $(A_t^1, \ldots, A_t^N) $. In a similar vein, for any collection of sets $(\cX^i)_{i \in \cN}$, we denote $\times_{i\in\cN} \cX^i$ by $\cX$. We denote the indicator function of a set $A$ by $I_{A}(\cdot)$.
For any finite set $\mathcal{S}$, $\Delta(\mathcal{S})$ represents the space of probability measures on $\mathcal{S}$ and $|\mathcal{S}|$ represents its cardinality. We denote by $\mathcal{P}^g$ (or $\E^g$) the probability measure generated by (or expectation with respect to) strategy profile $g$. We denote the set of real numbers by $\mathbb{R}$. For a probabilistic strategy profile of players $(\sigma_t^i)_{i\in \cN}$ where the probability of action $a_t^i$ conditioned on $(a_{1:t-1},x_{1:t}^i)$ is given by $\sigma_t^i(a_t^i|a_{1:t-1},x_{1:t}^i)$, we use the notation $\sigma_t^{-i}(a_t^{-i}|a_{1:t-1},x_{1:t}^{-i})$ to represent $\prod_{j\neq i} \sigma_t^j(a_t^j|a_{1:t-1},x_{1:t}^j)$.
All equalities/inequalities involving random variables are to be interpreted in the \emph{a.s.} sense.
For mappings with range function sets $f: \mathcal{A} \rightarrow (\mathcal{B} \rightarrow \mathcal{C})$ we use square brackets $f[a] \in\mathcal{B} \rightarrow\mathcal{C}$ to denote the image of $a\in\mathcal{A}$ through $f$ and parentheses $f[a](b) \in \mathcal{C}$ to denote the image of $b\in\mathcal{B}$ through $f[a]$.
A controlled Markov process with state $X_t$, action $A_t$, and horizon $\mathcal{T}$ is denoted by $(X_t,A_t)_{t\in\mathcal{T}}$.

The paper is organized as follows. In Section~\ref{sec:Model}, we present the model. In Section~\ref{sec:Sol_Concepts}, we present the solution concepts of perfect Bayesian and Perfect Stackelberg equilibrium.
In Section~\ref{sec:Result}, we present a two-step backward-forward recursive algorithm to construct a strategy profile and a sequence of beliefs of the dynamic game considered.  In Section~\ref{sec:Result2}, we extend that methodology to multiple receivers.

\section{Model}
\label{sec:Model}
Suppose there are two players, a sender and a receiver. Sender observes a controlled Markov process $\{X_t\}_t$ privately such that
\seq{
\eq{
P(x_t|x_{1:t-1}, a_{1:t-1}) &= Q(x_t|x_{t-1},a_{t-1} )
}
}
where $a_t\in\cA$ is the action taken by the receiver at time $t$. Sender takes action $s_t \in \cS$ at time $t$ upon observing $(s_{1:t-1}, a_{1:t-1})$, which is common information among players, and $x_{1:t}$, which is sender's private information. The sets $\cA, \cX,\cS$ are assumed to be finite and we also assume that kernel $Q_{x}$ has full support. Players play alternatively such that sender plays at odd times and the receiver plays at the even times. At the end of interval $t$, player $i$ receives an instantaneous reward $R^i(x_t,a_t)$.
All reward functions, priors and the update kernels are assumed to be common knowledge. We also assume that the receiver observes the rewards it receives $\{R_t^r\}_t$ during the course of the game. These could be understood as additional observation of the state by the receiver. We note that these rewards are a function of current state and action of the receiver, which the sender perfectly observes.
Let $g^i = (g^i_t)_t$ be a probabilistic strategy of player $i\in \{ S,R\}$ where $g^s_t : (\cS\times\cA\times\mathbb{R})^{t-1}\times (\cX)^{t} \to \Delta(\cA^s)$ and $g^r_t : (\cS\times\cA\times\mathbb{R})^{t-1} \to \Delta(\cA^r)$ such that player $i\in\{S,R\}$ plays action $a_t^i$ according to $A_t^s \sim g^s_t(\cdot|a_{1:t-1}, s_{1:t},r_{1:t-1}^r,x_{1:t})$ and $A_t^r \sim g^r_t(\cdot|a_{1:t-1}, r_{1:t-1}^r,s_{1:t})$. Let $g:= (g^i)_{i\in \{S,R\}}$ be a strategy profile of all players. The objective of the player $i$ is to maximize its total expected reward
\eq{ J^{i,g} := \E^g \left[ \sum_{t=1}^T R^i(X_t,A_t) \right] .
}

\subsection{Common agent approach}
Any history of this game at which players take action is of the form $h_t = (a_{1:t-1}, s_{1:t},r_{1:t-1}^r,x_{1:t})$. Let $\mathcal{H}_t$ be the set of such histories, $\mathcal{H}^T \defeq \cup_{t=0}^T \mathcal{H}_t $ be the set of all possible such histories in finite horizon and $\mathcal{H}^\infty \defeq \cup_{t=0}^\infty \mathcal{H}_t $ for infinite horizon. At any time $t$ the sender observes $h^s_t=h_t = (a_{1:t-1}, s_{1:t},r_{1:t-1}^r,x_{1:t})$ and both players together have $h^c_t = a_{1:t-1}, s_{1:t},r_{1:t-1}^r$ as common history. Since the receiver does not observe any private information, $h_t^r= h_t^c=(a_{1:t-1}, s_{1:t},r_{1:t-1}^r)$. Let $\mathcal{H}^i_t$ be the set of observed histories of player $i$ at time $t$ and $\mathcal{H}^c_t$ be the set of common histories at time $t$.

We recall that the sender and the receiver generate their actions at time $t$ as follows, $A_t^s \sim g^s_t(\cdot|a_{1:t-1}, s_{1:t},r_{1:t-1}^r,x_{1:t})$ and $A_t^r \sim g^r_t(\cdot|a_{1:t-1}, s_{1:t},r_{1:t-1}^r)$. 
An alternative way to view the problem is as follows. As is done in common information approach~\cite{NaMaTe13}, at odd time $t$, a fictitious common agent observes the common information $(a_{1:t-1},s_{1:t-2},r_{1:t-1}^r)$ and generates prescription function $\gamma_t^s= \psi_t^s[a_{1:t-1},s_{1:t-2},r_{1:t-1}^r]$. Sender uses its prescription function $\gamma_t^s$ to operate on its private information $x_{1:t}$ to produce its action $a_t^s$, $\gamma_t^s:\cX^t\to  \cP(\cA^s)$ and $a_t^s \sim\gamma_t^s(\cdot|x_{1:t})$. At even time $t$, the fictitious common agent observes the common information $(a_{1:t-2},s_{1:t-1},r_{1:t-1}^r)$ and generates prescription function $\gamma_t^r= \psi_t^r[a_{1:t-2},s_{1:t-1},r_{1:t-1}^r]$. Receiver uses its prescription function $\gamma_t^r$ to produce its action $a_t^r$, where $\gamma_t^r \in \cP(\cA^r)$ and $a_t^r \sim\gamma_t^r(\cdot)$. It is easy to see that for any $g$ policy profile of the players, there exists an equivalent $\psi$ profile of the common agent (and vice versa) that generates the same control actions for every realization of the information of the players.

Here, we will consider Markovian common agent's policy as follows. We call a common agent's policy be of ``type $\theta$" if the common agent observes the common beliefs $\mu_t$ at odd times and $\nu_t$ at even times derived from the common observation $a_{1:t-1},s_{1:t-2},r_{1:t-1}^r$ and $a_{1:t-2},s_{1:t-1}$, $r_{1:t-1}^r$ respectively, and generates prescription functions $\gamma_t^s = \theta_t[\mu_t]$ and $\gamma_t^r= \theta_t[\nu_t]$, where $\nu_t(x_t) = P^g(X_t = x_t|a_{1:t-1},s_{1:t-2},r_{1:t-1}^r)$ and $\mu_{t+1}(x_{t+1}) = P^g(X_{t+1}= x_{t+1}|a_{1:t-1},s_{1:t},r_{1:t-1}^r)$. Moreover, the sender's action only depends on its current private information $x_t$ i.e. $S_t\sim\gamma_t^s(\cdot|x_t)$.

In the next lemma we show that for any given $\theta$ policy, the belief states $\mu_t,\nu_t$ can be updated recursively as follows. Let  $\mu_1(x_1) := Q(x_1)$.

\begin{lemma} For any given policy of type $\theta$, there exists update functions $F,G$, independent of $\theta$, such that
\eq{
\nu_{t} &= F(\mu_t,\gamma_t^s,s_t)\\
\mu_{t+1} &= G(\nu_t,a_t,r_t^r) \label{eq:FG_update}
}
\end{lemma}
\begin{IEEEproof}
Please see Appendix~\ref{app:A}.
\end{IEEEproof}

\section{Solution Concepts}
\label{sec:Sol_Concepts}
\subsection{Solution Concept: Perfect Bayesian Equilibrium}
An appropriate concept of equilibrium for such games is PBE \cite{FuTi91book}, which consists of a pair $(\beta^*,\mu^*)$ of strategy profile $\beta^* = (\beta_t^{*,i})_{t \in \mathcal{T},i\in \cN}$ where $\beta_t^{*,i} : \mathcal{H}_t^i \to \Delta(\cA^i)$ and a belief profile $\mu^* = (^i\mu_t^{*})_{t \in \mathcal{T},i\in \cN}$ where $^i\mu_t^{*}: \mathcal{H}^i_t \to \Delta(\mathcal{H}_t)$ that satisfy sequential rationality so that for $ i =l,f,  t \in \mathcal{T},  h^{i}_t \in \mathcal{H}^i_t, {\beta^{i}}$
\begin{equation}
W_t^{i,\beta^{*,i},T}(h_t^i) \ge W_t^{i,\beta^i,T}(h_t^i)
\end{equation}
where the reward-to-go  is defined as
\begin{equation}
\hspace{-0.2cm}W_t^{i,\beta^i,T}(h_t^i) \triangleq \E^{{\beta}^{i} \beta^{*,-i},\, ^i\mu_t^*[h_t^i]}\left\{ \sum_{n=t}^T R_n^i(X_n, A_n)\big\lvert  h^i_t\right\}, \;\;   \label{eq:seqeq}
\end{equation}
and the beliefs are updated using Bayes' rule whenever possible. 
In general, a belief for player $i$ at time $t$, $^i\mu_t^{*}$ is defined on history $h_t = (a_{1:t-1}, s_{1:t},r_{1:t-1}^r, x_{1:t}) $ given its private history $h^i_t$. 
At any time $t$, the relevant uncertainty follower has is about the state history $x_{1:t} \in \times_{n=1}^t  \mathcal{X}$  and their future actions.
In our setting, we consider beliefs that are functions of each player's history $h^i_t$ only through the common history $h^c_t$ and are a belief on the current state only.
Here the follower's belief for each history $h^c_t=(a_{1:t-1}, s_{1:t},r_{1:t-1}^r)$ is derived from a common belief $\mu^*_t[a_{1:t-1}, s_{1:t},r_{1:t-1}^r]$. 
In order to anticipate followers actions through its strategy, the leader keeps track of this belief as well (and it can since it is derived from common information).
Thus we can sufficiently use the system of beliefs, $\mu^*=({\mu}^*_t)_{t\in\mathcal{T}}$, where $\mu^{*}_t: \mathcal{H}^c_t \to \Delta(\cX)$.

\subsection{Solution concept:Stackelberg Equilibrium}
\label{sec:Result_A}
An appropriate notion of equilibrium is Stackelberg equilibrium defined as follows. For a given strategy profile of the sender, $\sigma^s$, the receiver maximizes its total discounted expected utility over finite horizon $T$,
\eq{ 
\max_{\sigma^r}\E^{\sigma^s,\sigma^r} \left\{ \sum_{t=1}^T \delta^{t-1}R_t^r(X_t,A_t) \right\}.
}
Let $BR^r(\sigma^s)$ be the set of optimizing strategies of the receiver given a strategy $\sigma^s$ of the sender, i.e.
\eq{
BR^r(\sigma^s) = \arg\max_{\sigma^r} \E^{\sigma^s,\sigma^r} \left\{ \sum_{t=1}^T\delta^{t-1} R_t^r(X_t,A_t) \right\}
}
The sender finds its optimal strategy that maximizes its total expected discounted reward given that the receiver will use its best response to it,
\eq{ \tsigma^s \in\max_{ \sigma^s} \E^{\sigma^s,BR^r(\sigma^s)} \left\{ \sum_{t=1}^T\delta^{t-1} R_t^s(X_t,A_t) \right\},\label{eq:Stck_eq}
}
Then $(\tsigma^s,\tsigma^r)$ constitute a Stackelberg equilibrium where $\tsigma^r \in BR^r(\tsigma^s)$.

\subsection{Common Perfect Stackelberg equilibrium}
\label{sec:PBSE}
In this paper, we will consider sender' equilibrium policies that only depend on its current states $x_t$ and action history, i.e. at equilibrium, $a_t^s\sim \tsigma^s_t(\cdot|a_{1:t-1}, s_{1:t-1},r_{1:t-1}^r,x_t), a_t^r\sim \tsigma^s_t(\cdot|a_{1:t-1}, s_{1:t-1},r_{1:t-1}^r)$.\footnote{Note, however, that for the purpose of equilibrium, the optimization will be performed in the space of all possible strategies that may depend on the entire history of state.}

For the game considered, we introduce a notion of common Perfect Stackelberg Equilibrium (cPSE), inspired by perfect Bayesian equilibrium~\cite{FuTi91} as follows.

Let $(\tsigma,\mu,\nu)$ be a cPSE of the game, where $\mu = (\mu_t)_{t\in[T]},\nu = (\nu_t)_{t\in[T]}$, and for any $t$, $ (a_{1:t-2}, s_{1:t-1}$,$r_{1:t-1}^r$), $\mu_t[a_{1:t-2}, s_{1:t-1},r_{1:t-1}^r] \in\cP(\cX),\nu_t[a_{1:t-1}, s_{1:t-1},r_{1:t-1}^r] \in\cP(\cX)$ are the equilibrium belief on the current state $x_t$, given the action history $(a_{1:t-2}, s_{1:t-1},r_{1:t-1}^r),$\\$(a_{1:t-2}, s_{1:t-1},r_{1:t-1}^r)$ respectively, i.e. $\mu_t[a_{1:t-2}, s_{1:t-1},r_{1:t-1}^r](x_t) = P^{\tsigma}(x_t|a_{1:t-2}, s_{1:t-1},r_{1:t-1}^r)$, $ \nu_t[a_{1:t-1}, s_{1:t-1},r_{1:t-1}^r](x_t) = P^{\tsigma}(x_t|a_{1:t-1}, s_{1:t-1},r_{1:t-1}^r)$.
Then for all $t\in[T]$, \\$h_t^r=(a_{1:t-1}, s_{1:t-1},r_{1:t-1}^r),h_t^s=(a_{1:t-2}, s_{1:t-1},r_{1:t-1}^r)$, $x_{1:t}$, for any given $\sigma^s$, with some abuse of notation, let $BR^r(\sigma^s)$, which is best response of the receiver to any strategy $\sigma^l$ of the sender, be defined as, $\forall h_t^c$
\eq{
BR^r(\sigma^s) := \bigcap_t\bigcap_{h_t^c}\arg\max_{\sigma^r} \E^{\sigma^s,\sigma^r} \left\{ \sum_{n=t}^T \delta^{n-t}R_n^r(X_n,A_n) |h_t^c\right\}
}
and $\forall h_t^c$, let the set of optimum strategies of the sender be defined as,
\eq{ \tsigma^s \in\bigcap_t\bigcap_{h_t^c}  \arg\max_{ \sigma^s} \E^{\sigma^s,BR^r(\sigma^s)} \left\{ \sum_{n=t}^T \delta^{n-t}R_n^r(X_n,A_n) |h_t^c\right\},
}
Then $(\tsigma^s,\tsigma^r)$ constitute a cPSE of the game where $\tsigma^r \in BR_t^r(\tsigma^s)\ \forall \ t\in[T]$\footnote{Note that we condition on the common information and not on the actual observed histories of the players}.

\begin{definition}
We call a strategy profile $\sigma$ Markov cPSE if it is a cPSE of type $\theta$.
\end{definition}
In the next section, we design an algorithm to compute all Markovian cPSE of the game.

\section{Single receiver}
\label{sec:Result}
\subsection{PBE methodology}

In the following, we adapt the methodology presented in~\cite{VaSiAn19} to compute PBE of this game.

\subsection{Backward Recursion}
In this section, we define an equilibrium generating function $\theta=(\theta^i_t)_{i\in\{s,r\},t\in[T]}$, 
and a sequence of functions $(V_t^s,V_t^r)_{t\in \{ 1,2, \ldots T+1\}}$, 
in a backward recursive way, as follows. 
\begin{itemize}
\item[1.] Initialize $\forall \pi_{T+1}\in \mathcal{P}(\cX), x_{T+1}\in \cX$,
\eq{
V^r_{T+1}(\mu_{T+1}) &\defeq 0\\
V^{s+}_{T+1}(\mu_{T+1},x_{T+1}) &\defeq 0.   \label{eq:VT+1_1}
}
\item[2.] For $t = T,T-1, \ldots 1, \ \forall \nu_t\in \mathcal{P}(\cX)$, let $\tilde{\gamma}_t^r=\theta^r_t[\nu_t]$ be generated as follows. $\tilde{\gamma}_t^r$ is the solution of the following optimization. 
\eq{
\tgamma_t^r&\in  \arg\max_{\gamma^r_t} \E^{\gamma^r_t\,\nu_t} \left\{ R_t^r(X_t,A_t) +\delta V_{t+1}^r(G(\nu_t, A_t,R_t^r)) \big\lvert \nu_t \right\} , \label{eq:m_FP1_1}
}
where the expectation in~(\ref{eq:m_FP1_1}) is defined with respect to random variables $(X_t,A_t,R_t^r)$ through the measure $\nu_t(x_t){\gamma}^r_t(a_t)\mathbbm{1}(R_t^r = R_t^r(x_t,a_t))$,
Let 
\eq{
V^{s}_{t+1}(x_t,\nu_{t},x_{t+1}) &\defeq  \E^{\tgamma^r_t\,\nu_t} \left\{V^{s+}_{t+1}(G(\nu_t, A_t,R_t^r),x_{t+1})|x_t\right\}\\
V^{r+}_{t}(\nu_t) &\defeq  \;\E^{\tilde{\gamma}^{r}_t,\, \nu_t}\left\{ {R}_t^r (X_t,A_t) + \delta V_{t+1}^r(G(\nu_t, A_t,R_t^r)) \right\}
}
Let $\forall \nu_t\in \mathcal{P}(\cX), \tilde{\gamma}_t^s=\theta^s_t[\mu_t],$ be generated as follows. $\tilde{\gamma}_t^s$ is the solution of the following fixed-point equation.
\eq{
\tgamma_t^s(\cdot|x_t)&\in  \arg\max_{\gamma^s_t(\cdot|x_t)} \E^{\gamma^s_t(\cdot|x_t),\theta_t^r[F(\mu_t, \tilde{\gamma}^s_t, S_t)]\,\mu_t} \left\{ R_t^s(x_t,A_t) +\delta V_{t+1}^s(x_t,F(\mu_t, \tilde{\gamma}^s_t, S_t), X_{t+1}) \big\lvert \mu_t,x_t \right\} , \label{eq:m_FP2_1}
}
Let
\eq{
 V^{s+}_{t}(\mu_{t},x_t)& \defeq  \;\E^{\tilde{\gamma}^{s}_t(\cdot|x_t)\theta_t^r[F(\mu_t, \tilde{\gamma}^s_t, S_t)]\, \mu_{t}}\left\{ {R}_t^s (x_t,A_t) + \delta V_{t+1}^s (x_t,F(\mu_t, \tilde{\gamma}^s_t, S_t), X_{t+1})\big\lvert  x_t \right\}.  \label{eq:Vdef_1}\\
 V^{r}_{t}(\mu_t) &\defeq  \E^{\tilde{\gamma}^{s}_t,\, \mu_{t}}\left\{V^{r+}_{t}(F(\mu_t, \tilde{\gamma}^s_t, S_t)) \big\lvert\mu_t\right\}
}
where the expectation in \eqref{eq:m_FP2_1} is with respect to random variables $(S_t, A_t,X_{t+1})$ through the measure $\gamma^s_t(s_t|x_t) \theta_t^r[F(\mu_t, \tilde{\gamma}^s_t, S_t)](a_t)Q(x_{t+1}|x_t,a_t)$ and $F$ is defined in Appendix~\ref{app:A}. 



   \end{itemize}

\subsection{Forward Recursion}
Based on $\theta$ defined in the backward recursion above, we now construct a set of strategies $\tsigma$ (through beliefs $\mu$) in a forward recursive way as follows. 
\begin{itemize} 
\item[1.] Initialize at time $t=1$, 
\eq{
\mu_1[\phi](x_1) &:=  Q(x_1). \label{eq:mu*def0}
}
\item[2.] For $t =1,2 \ldots T, \forall i =1,2, a_{1:t}\in \cH_{t+1}^c, x_{1:t} \in\cX^t$
\eq{
\tsigma_{t}^{r}(a_{t}|a_{1:t-2},s_{1:t},r_{1:t-1}^r) &:= \theta_{t}^r[\nu_{t}[a_{1:t-2},s_{1:t},r_{1:t-1}^r]](a_{t}) \\
 \tsigma_{t}^{s}(s_{t}|a_{1:t-2},s_{1:t-1},r_{1:t-1}^r) &:= \theta_{t}^s[\mu_{t}[a_{1:t-2},s_{1:t-1},r_{1:t-1}^r]](s_{t}|x_{t})
 \label{eq:beta*def}
}
\vspace{-0.5cm}
\seq{
\eq{
\nu_{t}[a_{1:t-1},s_{1:t},r_{1:t-1}^r] &= F(\mu_t[a_{1:t-1},s_{1:t-1},r_{1:t-1}^r],\theta_t^s[\mu_t[a_{1:t-1},s_{1:t-1}]],s_t)\\
\mu_{t+1}[a_{1:t},s_{1:t-1},r_{1:t}^r] &= G(\nu_t[a_{1:t-1},s_{1:t-1},r_{1:t-1}^r],a_t,r_t^r)  \label{eq:mu*def}
}
}
\end{itemize}
where $F,G$ are defined in Appendix~\ref{app:A}.  
\begin{theorem}
\label{Thm:Main}
A strategy and belief profile $(\tsigma,\mu,\nu)$, as constructed through backward/forward recursion algorithm above is a PBE of the game.
\end{theorem}
\begin{IEEEproof}
Our model can be fit in the framework considered in~\cite{VaSiAn19} as follows.~\cite{VaSiAn19} considers a model where there are $N$ strategic players, each with a private type $x_t^i$ such that players' types are conditionally independent across players given the history of actions.
Although it assumes that all $N$ players act in all periods of the game, simultaneously, it can accommodate cases where at each time $t$, players are chosen through an exogenously defined Markov process. This is done by introducing a ``nature" player 0, who perfectly observes its state process $(X_t^0)_t\in\{S,R\}$, where the state process $X_t^0$ evolves as a deterministic process such that $x_t^0=S$ for odd times and $x_t^0=R$ for even times. Player 0 has reward function zero, and plays actions $a_t^0 = x_t^0$.
Once the quantity $a^0_{t-1}$ is publicly observed, all players can determine that the acting player (at time $t$) will be the one indicated by $a^0_{t-1}$ such that $a_{t-1}^0=x_{t-1}^0=S$ indicates sender plays in the game and $a_{t-1}^0=x_{t-1}^0=R$ indicates that the receiver plays in the game. This is achieved by setting $\forall i$, $R^i_t(x_t,a_t) = 0$ if $i \neq a_t^0$, and $Q_x(x_{t+1}^i|x_t^i,a_t) = Q_x(x_{t+1}^i|x_t^i,a_t^{a_t^0})$. Here, in each period only one player (player $a^0_t=  x_t^0$) acts in the game while all other non-acting players receive zero rewards during that period.

Then the above methodology can be seen as adaptation of the methodology considered in~\cite{VaSiAn19} to compute PBE of the game and result of the theorem is implied by~\cite[Theorem~1]{VaSiAn19}.
\end{IEEEproof}

\subsection{Stackelberg methodology: Single Receiver}

In the following, we adapt the methodology presented in~\cite{Va20} to compute cPSE of this game.

\subsection{Backward Recursion}
In this section, we define an equilibrium generating function $\theta=(\theta^i_t)_{i\in\{l,f\},t\in[T]}$, where $\theta^s_t : \mathcal{P}(\cX)\to \left\{\cX \to \mathcal{P}(\cA) \right\}, \theta^r_t : \mathcal{P}(\cX)\to\mathcal{P}(\cA)$ and a sequence of functions $(V_t^s,V_t^r)_{t\in \{ 1,2, \ldots T+1\}}$, where $V_t^s:  \mathcal{P}(\cX)\times \cX \to \mathbb{R}, V_t^r:  \mathcal{P}(\cX) \to \mathbb{R}$, in a backward recursive way, as follows.
\begin{itemize}
\item[1.] Initialize $\forall \pi_{T+1}\in \mathcal{P}(\cX), x_{T+1}\in \cX$,
\eq{
V^r_{T+1}(\mu_{T+1}) &\defeq 0\\
V^{s+}_{T+1}(\mu_{T+1}) &\defeq 0.   \label{eq:VT+1}
}
\item[2.] For $t = T,T-1, \ldots 1, \ \forall \nu_t\in \mathcal{P}(\cX)$, let $\tilde{\gamma}_t^r=\theta^r_t[\nu_t]$ be generated as follows. $\tilde{\gamma}_t^r$ is the solution of the following optimization. 
\eq{
\tgamma_t^r&\in  \arg\max_{\gamma^r_t} \E^{\gamma^r_t\,\nu_t} \left\{ R_t^r(X_t,A_t) +\delta V_{t+1}^r(G(\nu_t, A_t,R_t^r)) \big\lvert \nu_t \right\} , \label{eq:m_FP1a}
}
where the expectation in~(\ref{eq:m_FP1a}) is defined with respect to random variables $(X_t,A_t,R_t^r)$ through the measure $\nu_t(x_t){\gamma}^r_t(a_t)\mathbbm{1}(R_t^r = R_t^r(x_t,a_t))$,
Let 
\eq{
V^{s}_{t+1}(x_t,\nu_{t}) &\defeq  \E^{\tgamma^r_t\,\nu_t} \left\{V^{s+}_{t+1}(G(\nu_t, A_t,R_t^r))\right\}\\
V^{r+}_{t}(\nu_t) &\defeq  \;\E^{\tilde{\gamma}^{r}_t,\, \nu_t}\left\{ {R}_t^r (X_t,A_t) + \delta V_{t+1}^r(G(\nu_t, A_t,R_t^r)) \right\}
}
Let $\forall \nu_t\in \mathcal{P}(\cX), \tilde{\gamma}_t^s=\theta^s_t[\mu_t],$ be generated as follows. $\tilde{\gamma}_t^s$ is the solution of the following fixed-point equation.
\eq{
\tgamma_t^s&\in  \arg\max_{\gamma^s_t} \E^{\gamma^s_t,\theta_t^r[F(\mu_t, {\gamma}^s_t, S_t)]\,\mu_t} \left\{ R_t^s(X_t,A_t) +\delta V_{t+1}^s(X_t,F(\mu_t,{\gamma}^s_t, S_t)) \right\} \label{eq:m_FP2a}
}
Let
\eq{
 V^{s+}_{t}(\mu_{t})& \defeq  \;\E^{\tilde{\gamma}^{s}_t(\cdot|x_t)\theta_t^r[F(\mu_t, \tilde{\gamma}^s_t, S_t)]\, \mu_{t}}\left\{ {R}_t^s (X_t,A_t) + \delta V_{t+1}^s (X_t,F(\mu_t, \tilde{\gamma}^s_t, S_t), X_{t+1}) \right\}.  \label{eq:Vdef}\\
 V^{r}_{t}(\mu_t) &\defeq  \E^{\tilde{\gamma}^{s}_t,\, \mu_{t}}\left\{V^{r+}_{t}(F(\mu_t, \tilde{\gamma}^s_t, S_t)) \right\}
}
where the expectation in \eqref{eq:m_FP2a} is with respect to random variables $(X_t,S_t, A_t)$ through the measure $\mu_t(x_t)\gamma^s_t(s_t|x_t) \theta_t^r[F(\mu_t,{\gamma}^s_t, S_t)](a_t)Q(x_{t+1}|x_t,a_t)$ and $F$ is defined in Appendix~\ref{app:A}. 

   \end{itemize}

\subsection{Forward Recursion}
Based on $\theta$ defined in the backward recursion above, we now construct a set of strategies $\tsigma$ (through beliefs $\mu$) in a forward recursive way as follows. 
\begin{itemize} 
\item[1.] Initialize at time $t=1$, 
\eq{
\mu_1[\phi](x_1) &:=  Q(x_1). \label{eq:mu*def0}
}
\item[2.] For $t =1,2 \ldots T, \forall i =1,2, a_{1:t}\in \cH_{t+1}^c, x_{1:t} \in\cX^t$
\eq{
\tsigma_{t}^{r}(a_{t}|a_{1:t-1}, s_{1:t-1},r_{1:t-1}^r) &:= \theta_{t}^r[\nu_{t}[a_{1:t-1}, s_{1:t-1},r_{1:t-1}^r]](a_{t}) \\
 \tsigma_{t}^{s}(s_{t}|a_{1:t-2}, s_{1:t-1},r_{1:t-1}^r,x_{1:t}) &:= \theta_{t}^s[\mu_{t}[a_{1:t-2}, s_{1:t-1},r_{1:t-1}^r]](s_{t}|x_{t})
 \label{eq:beta*def}
}
\vspace{-0.5cm}
\seq{
\eq{
\nu_{t}[a_{1:t-1}, s_{1:t},r_{1:t-1}^r] &= F(\mu_t[a_{1:t-1}, s_{1:t-1},r_{1:t-1}^r],\theta_t^s[\mu_t[a_{1:t-1}, s_{1:t-1},r_{1:t-1}^r]],s_{t})\\
\mu_{t+1}[a_{1:t}, s_{1:t},r_{1:t}^r] &= G(\nu_t[a_{1:t-1}, s_{1:t},r_{1:t-1}^r],a_t,r_t^r)  \label{eq:mu*def}
}
}
\end{itemize}
where $F,G$ are defined in Appendix~\ref{app:A}.  
\begin{theorem}
\label{Thm:Main}
A strategy and belief profile $(\tsigma,\mu,\nu)$, as constructed through backward/forward recursion algorithm above is a cPSE of the game.
\end{theorem}
\begin{IEEEproof}

Then the above methodology can be seen as adaptation of the methodology considered in~\cite{Va20} to compute cPSE of the game and result of the theorem is implied by~\cite[Theorem~1]{Va20}. In this case, since only the sender has private information and thus can influence the beliefs, it solves for the optimization problem in~\eqref{eq:m_FP2a}. The receiver however doesn't have any private information and solves for an optimization problem in~\eqref{eq:m_FP1a}.
\end{IEEEproof}

\section{Multiple players}
\label{sec:Result2}
In this section, we consider the case when there are multiple receivers. As before, each receiver $i$ takes action at time $t$, $A_t^{r,i}$ as $A_t^{r,i}\sim g^{r,i}_t(\cdot|a_{1:t-2},s_{1:t-1},r_{1:t-1}^r)$. The above methodologies can be extended as follows,
\subsection{PBE methodology: Multiple Receivers}
\subsection{Backward Recursion}
In this section, we define an equilibrium generating function $\theta=(\theta^i_t)_{i\in\{s,r^1,\ldots,r^N\},t\in[T]}$, 
and a sequence of functions $(V_t^s,V_t^{r,1},\ldots, V_t^{r,N})_{t\in \{ 1,2, \ldots T+1\}}$, 
in a backward recursive way, as follows. 
\begin{itemize}
\item[1.] Initialize $\forall \pi_{T+1}\in \mathcal{P}(\cX), x_{T+1}\in \cX$,
\eq{
V^{r,i}_{T+1}(\mu_{T+1}) &\defeq 0\\
V^{s+}_{T+1}(\mu_{T+1},x_{T+1}) &\defeq 0.   \label{eq:VT+1}
}
\item[2.] For $t = T,T-1, \ldots 1, \ \forall \nu_t\in \mathcal{P}(\cX)$, let $\tilde{\gamma}_t^{r,i}=\theta^r_t[\nu_t]$ be generated as follows. $\tilde{\gamma}_t^{r,i}$ is the solution of the following fixed-point equation. 
\eq{
\tgamma_t^{r,i}&\in  \arg\max_{\gamma^{r,i}_t} \E^{\gamma^{r,i}_t,\tgamma^{r,-i}_t\,\nu_t} \left\{ R_t^{r,i}(X_t,A_t) +\delta V_{t+1}^{r,i}(G(\nu_t, A_t,R_t^r)) \big\lvert \nu_t \right\} , \label{eq:m_FP1}
}
where the expectation in~(\ref{eq:m_FP1}) is defined with respect to random variables $(X_t,A_t)$ through the measure $\nu_t(x_t){\gamma}^{r,i}_t(a^i_t){\tgamma}^{r,-i}_t(a^{-i}_t)$.
Let 
\eq{
V^{s}_{t+1}(x_t,\nu_{t},x_{t+1}) &\defeq  \E^{\tgamma^r_t(\cdot)\,\nu_t} \left\{V^{s+}_{t+1}(G(\nu_t, A_t,R_t^r),x_{t+1})|x_t\right\}\\
V^{r+,i}_{t}(\nu_t) &\defeq  \;\E^{\tilde{\gamma}^{r}_t,\, \nu_t}\left\{ {R}_t^{r,i} (X_t,A_t) + \delta V_{t+1}^{r,i}(G(\nu_t, A_t,R_t^r) \right\}
}
Let $\forall \nu_t\in \mathcal{P}(\cX), \tilde{\gamma}_t^s=\theta^s_t[\mu_t],$ be generated as follows. $\tilde{\gamma}_t^s$ is the solution of the following fixed-point equation.
\eq{
\tgamma_t^s&\in  \arg\max_{\gamma^s_t(\cdot|x_t)} \E^{\gamma^s_t(\cdot|x_t),\theta_t^r[F(\mu_t, \tilde{\gamma}^s_t, S_t)]\,\mu_t} \left\{ R_t^s(x_t,A_t) +\delta V_{t+1}^s(x_t,F(\mu_t, \tilde{\gamma}^s_t, S_t), X_{t+1}) \big\lvert \mu_t,x_t \right\} , \label{eq:m_FP2}
}
Let
\eq{
 V^{s+}_{t}(\mu_{t},x_t)& \defeq  \;\E^{\tilde{\gamma}^{s}_t(\cdot|x_t)\theta_t^r[F(\mu_t, \tilde{\gamma}^s_t, S_t)]\, \mu_{t}}\left\{ {R}_t^s (x_t,A_t) + \delta V_{t+1}^s (x_t,F(\mu_t, \tilde{\gamma}^s_t, S_t), X_{t+1})\big\lvert  x_t \right\}.  \label{eq:Vdef}\\
 V^{r,i}_{t}(\mu_t) &\defeq  \E^{\tilde{\gamma}^{s}_t,\, \mu_{t}}\left\{V^{r+,i}_{t}(F(\mu_t, \tilde{\gamma}^s_t, S_t)) \big\lvert\mu_t\right\}
}
where the expectation in \eqref{eq:m_FP2} is with respect to random variables $(S_t, A_t,X_{t+1})$ through the measure $\gamma^s_t(s_t|x_t) \theta_t^r[F(\mu_t, \tilde{\gamma}^s_t, S_t)](a_t)Q(x_{t+1}|x_t,a_t)$ and $F$ is defined in Appendix~\ref{app:A}. 



   \end{itemize}

\subsection{Forward Recursion}
Based on $\theta$ defined in the backward recursion above, we now construct a set of strategies $\tsigma$ (through beliefs $\mu$) in a forward recursive way as follows. 
\begin{itemize} 
\item[1.] Initialize at time $t=1$, 
\eq{
\mu_1[\phi](x_1) &:=  Q(x_1). \label{eq:mu*def0}
}
\item[2.] For $t =1,2 \ldots T, \forall i =1,2, a_{1:t}\in \cH_{t+1}^c, x_{1:t} \in\cX^t$
\eq{
\tsigma_{t}^{r,i}(a_{t}|a_{1:t-1},s_{1:t},r_{1:t-1}^r) &:= \theta_{t}^{r,i}[\nu_{t}[a_{1:t-1},s_{1:t-1},r_{1:t-1}^r]](a^i_{t}) \\
 \tsigma_{t}^{s}(s_{t}|a_{1:t-1},s_{1:t-1},r_{1:t-1}^r,x_{1:t}) &:= \theta_{t}^s[\mu_{t}[a_{1:t-1},s_{1:t-1},r_{1:t-1}^r]](s_{t}|x_{t})
 \label{eq:beta*def}\\
\nu_{t}[a_{1:t-1},s_{1:t},r_{1:t-1}^r] &= F(\mu_t[a_{1:t-1},s_{1:t-1},r_{1:t-1}^r],\theta_t^s[\mu_t[a_{1:t-1},s_{1:t-1},r_{1:t-1}^r]],s_t)\\
\mu_{t+1}[a_{1:t},s_{1:t-1},r_{1:t}^r] &= G(\nu_t[a_{1:t-1},s_{1:t-1},r_{1:t-1}^r],a_t,r_t^r)  \label{eq:mu*def}
}

\end{itemize}
where $F,G$ are defined in Appendix~\ref{app:A}.  
\begin{theorem}
\label{Thm:Main}
A strategy and belief profile $(\tsigma,\mu,\nu)$, as constructed through backward/forward recursion algorithm above is a PBE of the game.
\end{theorem}
\begin{IEEEproof}
By similar arguments as that in Theorem~1 above, the result is implied by~\cite[Theorem~1]{VaSiAn19}.
\end{IEEEproof}

\subsection{cPSE methodology: Multiple Receivers}
In this section, we consider the case when there are multiple receivers. As before, each receiver $i$ takes action at time $t$, $A_t^{r,i}$ as $A_t^{r,i}\sim g^{r,i}_t(\cdot|a_{1:t-1}, s_{1:t-1},r_{1:t-1}^r)$. The above methodology can be extended as follows.
\subsection{Backward Recursion}
In this section, we define an equilibrium generating function $\theta=(\theta^i_t)_{i\in\{s,r^1,\ldots,r^N\},t\in[T]}$, 
and a sequence of functions $(V_t^s,V_t^{r,1},\ldots, V_t^{r,N})_{t\in \{ 1,2, \ldots T+1\}}$, 
in a backward recursive way, as follows. 
\begin{itemize}
\item[1.] Initialize $\forall \pi_{T+1}\in \mathcal{P}(\cX), x_{T+1}\in \cX$,
\eq{
V^{r,i}_{T+1}(\mu_{T+1}) &\defeq 0\\
V^{s+}_{T+1}(\mu_{T+1}) &\defeq 0.   \label{eq:VT+1}
}
\item[2.] For $t = T,T-1, \ldots 1, \ \forall \nu_t\in \mathcal{P}(\cX)$, for all $i$, let $\tilde{\gamma}_t^{r,i}=\theta^{r,i}_t[\nu_t]$ be generated as follows. $\tilde{\gamma}_t^{r,i}$ is the solution of the following fixed-point equation. 
\eq{
\tgamma_t^{r,i}&\in  \arg\max_{\gamma^{r,i}_t} \E^{\gamma^{r,i}_t,\tgamma^{r,-i}_t\,\nu_t} \left\{ R_t^{r,i}(X_t,A_t) +\delta V_{t+1}^{r,i}(G(\nu_t, A_t,R_t^r)) \big\lvert \nu_t \right\} , \label{eq:m_FP1b}
}
where the expectation in~(\ref{eq:m_FP1b}) is defined with respect to random variables $(X_t,A_t)$ through the measure $\nu_t(x_t){\gamma}^{r,i}_t(a^i_t){\tgamma}^{r,-i}_t(a^{-i}_t)$. Note that the above equation is similar to a fixed-point equation corresponding to that of a Bayesian Nash.
Let 
\eq{
V^{s}_{t+1}(x_t,\nu_{t}) &\defeq  \E^{\tgamma^r_t\,\nu_t} \left\{V^{s+}_{t+1}(G(\nu_t, A_t,R_t^r)\right\}\\
V^{r+,i}_{t}(\nu_t) &\defeq  \;\E^{\tilde{\gamma}^{r}_t,\, \nu_t}\left\{ {R}_t^{r,i} (X_t,A_t) + \delta V_{t+1}^{r,i}(G(\nu_t, A_t,R_t^r) \right\}
}
Let $\forall \nu_t\in \mathcal{P}(\cX), \tilde{\gamma}_t^s=\theta^s_t[\mu_t],$ be generated as follows. $\tilde{\gamma}_t^s$ is the solution of the following fixed-point equation.
\eq{
\tgamma_t^s&\in  \arg\max_{\gamma^s_t} \E^{\gamma^s_t,\theta_t^r[F(\mu_t, {\gamma}^s_t, S_t)]\,\mu_t} \left\{ R_t^s(X_t,A_t) +\delta V_{t+1}^s(X_t,F(\mu_t, {\gamma}^s_t, S_t)) \big\lvert \mu_t\right\} , \label{eq:m_FP2b}
}
Let
\eq{
 V^{s+}_{t}(\mu_{t})& \defeq  \;\E^{\tilde{\gamma}^{s}_t\theta_t^r[F(\mu_t, \tilde{\gamma}^s_t, S_t)]\, \mu_{t}}\left\{ {R}_t^s (X_t,A_t) + \delta V_{t+1}^s (X_t,F(\mu_t, \tilde{\gamma}^s_t, S_t))\right\}.  \label{eq:Vdefb}\\
 V^{r,i}_{t}(\mu_t) &\defeq  \E^{\tilde{\gamma}^{s}_t,\, \mu_{t}}\left\{V^{r+,i}_{t}(F(\mu_t, \tilde{\gamma}^s_t, S_t)) \big\lvert\mu_t\right\}
}
where the expectation in \eqref{eq:m_FP2b} is with respect to random variables $(X_t,S_t, A_t)$ through the measure $\pi_t(x_t)\gamma^s_t(s_t|x_t) \theta_t^r[F(\mu_t, {\gamma}^s_t, S_t)](a_t)Q(x_{t+1}|x_t,a_t)$ and $F$ is defined in Appendix~\ref{app:A}. 



   \end{itemize}

\subsection{Forward Recursion}
Based on $\theta$ defined in the backward recursion above, we now construct a set of strategies $\tsigma$ (through beliefs $\mu$) in a forward recursive way as follows. 
\begin{itemize} 
\item[1.] Initialize at time $t=1$, 
\eq{
\mu_1[\phi](x_1) &:=  Q(x_1). \label{eq:mu*def0}
}
\item[2.] For $t =1,2 \ldots T, \forall i =1,2, a_{1:t}\in \cH_{t+1}^c, x_{1:t} \in\cX^t$
\eq{
\tsigma_{t}^{r}(a_{t}|a_{1:t-1}, s_{1:t-1},r_{1:t-1}^r) &:= \theta_{t}^r[\nu_{t}[a_{1:t-1}, s_{1:t-1},r_{1:t-1}^r]](a_{t}) \\
 \tsigma_{t}^{s}(s_{t}|a_{1:t-2}, s_{1:t-1},r_{1:t-1}^r,x_{1:t}) &:= \theta_{t}^s[\mu_{t}[a_{1:t-2}, s_{1:t-1},r_{1:t-1}^r]](s_{t}|x_{t})
 \label{eq:beta*def}
}
\vspace{-0.5cm}
\seq{
\eq{
\nu_{t}[a_{1:t-1}, s_{1:t},r_{1:t-1}^r] &= F(\mu_t[a_{1:t-1}, s_{1:t-1},r_{1:t-1}^r],\theta_t^s[\mu_t[a_{1:t-1}, s_{1:t-1},r_{1:t-1}^r]],s_{t})\\
\mu_{t+1}[a_{1:t}, s_{1:t},r_{1:t}^r] &= G(\nu_t[a_{1:t-1}, s_{1:t},r_{1:t-1}^r],a_t,r_t^r)  \label{eq:mu*def}
}
}

\end{itemize}
where $F,G$ are defined in Appendix~\ref{app:A}.  
\begin{theorem}
\label{Thm:Main}
A strategy and belief profile $(\tsigma,\mu,\nu)$, as constructed through backward/forward recursion algorithm above is a cPSE of the game.
\end{theorem}
\begin{IEEEproof}
The result is implied by~\cite[Theorem~1]{Va20} using similar arguments as in the proof of Theorem~1 above. Note that in this case, there are multiple receivers maximizing their rewards in each step in~\eqref{eq:m_FP1b}, thus instead of an optimization problem, they are solving a fixed-point equation similar to that of a Bayesian Nash equilibrium.
\end{IEEEproof}
\begin{appendix}
\section{}
\label{app:A}
\begin{IEEEproof}
\eq{
\nu_t(x_{t})& = P^{\theta}(x_{t}|s_{1:t},a_{1:t-1},r_{1:t-1}^r)\\
&=\frac{P^{\theta}(x_t,s_t|s_{1:t-1},a_{1:t-1},r_{1:t-1}^r)}{\sum_{x_t}P^{\theta}(x_t,s_t|s_{1:t-1},a_{1:t-1},r_{1:t-1}^r)}\\
&=\frac{\mu_t(x_t)\gamma_t^s(s_t|x_t)}{\sum_{x_t}\mu_t(x_t)\gamma_t^s(s_t|x_t)}.
}

Thus,
\eq{
\nu_t = F(\mu_t,\gamma_t^s,s_t) 
}

\eq{
\mu_{t+1}(x_{t+1})& = P^{\theta}(x_{t+1}|s_{1:t},a_{1:t},r_{1:t}^r)\\
&=\frac{\sum_{x_t}P^{\theta}(x_t,a_t,r_t^r,x_{t+1}|s_{1:t},a_{1:t-1},r_{1:t-1}^r)}{\sum_{x_t}P^{\theta}(x_t,a_t,r_t^r|s_{1:t},a_{1:t-1},r_{1:t-1}^r)}\\
&=\frac{\sum_{x_t}\nu_t(x_t)\gamma_t^r(a_t)\mathbbm{1}(r_t^r = R_t^r(x_t,a_t))Q_x(x_{t+1}|x_t,a_t)}{\sum_{x_t}\nu_t(x_t)\gamma_t^r(a_t)\mathbbm{1}(r_t^r = R_t^r(x_t,a_t))}\\
&=\frac{\sum_{x_t}\nu_t(x_t)\mathbbm{1}(r_t^r = R_t^r(x_t,a_t))Q_x(x_{t+1}|x_t,a_t)}{\sum_{x_t}\nu_t(x_t)\mathbbm{1}(r_t^r = R_t^r(x_t,a_t))}
}

Thus,
\eq{
\mu_{t+1} = G(\nu_t,a_t,r_t^r) 
}

\end{IEEEproof}

\end{appendix}
\bibliographystyle{IEEEtran}

	
        \end{document}